\documentclass[twocolumn]{jpsj2}
\usepackage{amsfonts,amssymb}
\usepackage[mathscr]{euscript}

\newcommand*{\be}{\begin{equation}}
\newcommand*{\ee}{\end{equation}}
\newcommand*{\ba}{\begin{array}}
\newcommand*{\ea}{\end{array}}
\newcommand*{\bea}{\begin{eqnarray}}
\newcommand*{\eea}{\end{eqnarray}}
\newcommand*{\bean}{\begin{eqnarray*}}
\newcommand*{\eean}{\end{eqnarray*}}

\newcommand*{\lp}{\left(}
\newcommand*{\rp}{\right)}

\newcommand*{\lc}{\left\{}
\newcommand*{\rc}{\right\}}

\newcommand*{\la}{\langle}
\newcommand*{\La}{\left\la}
\newcommand*{\ra}{\rangle}
\newcommand*{\Ra}{\right\ra}

\renewcommand*{\d}{\textrm{d}}
\newcommand*{\g}{\mbox{\slshape g}}

\newcommand*{\CMP}{Commun. Math. Phys.\ }
\newcommand*{\CMPL}{Cond. Matt. Phys.\ }
\newcommand*{\JMP}{J. Math. Phys.\ }
\newcommand*{\JPA}{J. Phys. A\ }
\newcommand*{\JPSJ}{J. Phys. Soc. Japan\ }
\newcommand*{\PLA}{Phys. Lett. A\ }
\newcommand*{\PTP}{Progr. Theor. Phys.\ }
\newcommand*{\PRev}{Phys. Rev.\ }
\newcommand*{\RMaP}{Rev. Math. Phys.\ }
\newcommand*{\RMoP}{Rev. Mod. Phys.\ }
\newcommand*{\ZP}{Z. Phys.\ }

\newcommand*{\venus}{{\scriptscriptstyle+}\hspace{-1.5mm}^{\circ}}
\newcommand*{\ti}[1]{{#1}}

\makeatletter
\def\@typeset{}
\makeatother

\title{Boson and Fermion Brownian Motion%
}

\author
{Alexander E. \textsc{Kobryn}\thanks{%
Present address: Institute for Molecular Science,
Myodaiji, Okazaki, Aichi 444-8585, Japan},
Tsuyoshi \textsc{Hayashi} and Toshihico \textsc{Arimitsu}}

\inst{Institute of Physics, University of Tsukuba, Ibaraki 305-8571, Japan}

\recdate{February 26, 2003; Accepted April 17, 2003}

\abst{Dynamics of quantum systems which are
perturbed by linear coupling to the reservoir stochastically can
be studied in terms of quantum stochastic differential equations
(for example, quantum stochastic Liouville equation and quantum
Langevin equation). To work it out one needs definition of quantum
Brownian motion. Since till very recent times only its boson
version has been known, in present paper we demonstrate definition which
makes possible consideration also for fermion Brownian motion.}

\kword{quantum Brownian motion, quantum stochastic differential equations}

\begin{document}

\maketitle

Description of time-dependent behavior of non-equi\-li\-bri\-um
systems involving stochastic forces can be given with the help of
the Langevin equation. It is the stochastic differential equation
for dynamical variables and is of the fundamental importance in
the theory of Brownian motion \cite{Chandrasekhar43,Wang45}.
Random forces in the Langevin equation are usually described by
Gaussian white stochastic processes \cite{Mazo02} since such
description is a convenient model for processes with short
correlation times. Similarly, from the mathematical point of view,
white noise can be considered to be a limit of some well specified
stochastic processes. Stochastic integral with respect to such
processes is defined as a kind of the Riemann--Stieltjes one
\cite{Gardiner85book} where multiplication between the stochastic
increment and integrand is commonly considered in the form of
It\^o~\cite{Ito44} or Stratonovich~\cite{Stratonovich66}.

The Langevin equation can be used to calculate various time
correlation functions. Now it is radically extended to solve
numerous problems arising in different areas
\cite{Shuler69,Coffey85,Sobczyk91,vanKampen92,Coffey96}. In
particular, the theory of Brownian motion itself has been extended
to situations where the ``Brownian particle'' is not a real
particle anymore, but instead some collective properties of a
macroscopic system. Corresponding equation in the phase space or
the Liouville space of quantum mechanical statistical operators
can be also considered as a sort of stochastic differential
equation. In order to investigate classical stochastic systems,
the stochastic Liouville equation was introduced first by Anderson
\cite{Anderson54} and Kubo \cite{Kubo54,Kubo62,Kubo63}.

There were several attempts to extend the classical theory (both
Langevin and stochastic Liouville equations) for quantum cases.
Study of the Langevin equation for quantum systems has its origin
in papers by Senitzky \cite{Senitzky60},
Schwinger \cite{Schwinger61}, Haken
\cite{Haken64,Haken70} and Lax \cite{Lax66}, where
they investigated a quantum mechanical damped harmonic oscillator
in connection with laser systems. In particular, it was shown that
the quantum noise, i.e.~the spontaneous emission, can be treated
in a way similar to the thermal fluctuations, and that the noise
source has non-zero second moments proportional to a quantity
which can be associated with a quantum analogue of a diffusion
coefficient. As it was noticed by Kubo \cite{Kubo69} in his
discussion with van~Kampen, the random force must be an operator
defined in its own Hilbert space, which does not happen in
classical case since there is no consideration of space for the
random force.

Mathematical study of the quantum stochastic processes was
initiated by Davies \cite{Davies69,Davies76}. At present,
essential progress in this subject is due to the massive invasion
of mathematicians in recent years
\cite{Cockroft77,Hudson81,Hudson84a,Hudson84c,%
Accardi82,Accardi02,Parthasarathy85,Parthasarathy92}. For
instance, quantum mechanical analogues of Wiener processes
\cite{Cockroft77} and quantum It\^o formula for boson systems
\cite{Hudson81,Hudson84a,Hudson84c} were defined first by Hudson
and co-authors. The classical Brownian motion is replaced here by
the pair of one-parameter unitary group authomorphisms, namely by
the annihilation and creation boson random force operators with
time indices in the boson Fock space, named quantum Brownian
motion. Fermion stochastic calculus was suggested first by
Applebaum, Hudson and Parthasarathy
\cite{Applebaum84a,Applebaum86,Hudson86,Parthasarathy86}. In these
papers, they de\-ve\-lo\-ped the fermion analog of the corresponding
boson theory \cite{Hudson84a} in which the annihilation and
creation processes are fermion field operators in the fermion Fock
space. Within the frame of this formalism, the It\^o--Clifford
integral \cite{Barnett82}---fermion analog of the classical
Brownian motion---is contained as a special case.

In this paper we give a definition of boson and fermion Brownian
motion following the approach suggested by Hudson and
Parthasarathy \cite{Hudson84a,Hudson86}. We do not pay much
attention to purely mathematical aspects of the subject. Instead
of that we try to represent the material in language which is easy
to understand for physicists. Also we consider quantum Brownian
motion with allowance for thermal degree of freedom.

Let $\Gamma^0_{\mathrm{s}}$ denotes the boson Fock space (the
symmetric Fock space) over the Hilbert space
$\mathscr{H}=L^2(\mathbb{R}_+)$ of square integrable functions,
and $b_t$ and $b_t^\dag$ denote, respectively, boson annihilation
and creation operators at time $t\in[0,\infty)$ satisfying the
canonical commutation relations
\begin{eqnarray}
\label{boson-commutators}
[b_t,b^\dag_s]=\delta(t-s),\quad
[b_t,b_s]=[b^\dag_t,b^\dag_s]=0.
\end{eqnarray}
The bra- and ket-vacuums $(|$ and $|)$, respectively, are defined by
\begin{eqnarray}
(|b^\dag_t=0,\quad
b_t|)=0.\label{b2}
\end{eqnarray}
The space $\Gamma^0_{\mathrm{s}}$ is equipped with a total family
of exponential vectors
\begin{subequations}
\label{exp-vectors}
\begin{align}
(e(f)|&=(|\exp\lc\int_0^\infty\d t\,f^*(t)b_t\rc,\\
|e(\g))&=\exp\lc\int_0^\infty\d t\,\g(t)b^\dag_t\rc|),
\end{align}
\end{subequations}
whose overlapping is
$(e(f)|e(\g))=\exp\!\lc\!\int_0^\infty\d t\,f^*(t)\g(t)\rc$.
Here $f,\g\in\mathscr{H}$ are elements of the set
$L^2(\mathbb{R}_+)$ of square integrable functions satisfying
$\int_0^\infty\d t\,|f(t)|^2<\infty$ and $\int_0^\infty\d
t\,|\g(t)|^2<\infty$. The dense span of exponential vectors we
denote by $\mathscr{E}$. Operators $b_t$, $b^\dag_t$ and
exponential vectors are characterized by the relations
\begin{eqnarray}
\label{d5}
(e(f)|b^\dag_t=(e(f)|f^*(t),\quad
b_t|e(\g))=\g(t)|e(\g)).
\end{eqnarray}

Let us introduce an operator $U_t$ defined as
\begin{eqnarray}
U_t=\sigma_<P_{[0,t]}+\sigma_>P_{(t,\infty)},
\end{eqnarray}
where $\sigma_<$ and $\sigma_>$ are two independent parameters
taking values $\pm1$, and $P_{[a,b]}$ ($a$$\leqslant$$b$) is an
operator on $\mathscr{H}$ of multiplication by the indicator
function whose action reads
\be
P_{[a,b]}\int_0^\infty\d t\,\g(t)
=\int_0^\infty\d t\,\theta(t-a)\theta(b-t)\g(t).\label{Pab}
\ee
Here, $\theta(t)$ is the step function specified as
\begin{eqnarray}
\theta(t)=\lc
\ba{cl}
1&\mbox{for}\ t\geqslant0,\\
0&\mbox{for}\ t<0.
\ea
\right.
\end{eqnarray}
Operator $P_{[a,b]}$ has the following properties:
\begin{equation}
P_{[a,b]}^2=P_{[a,b]},\!\quad
P_{[a,b]}^\dag=P_{[a,b]},\!\quad
P_{[a,b]}P_{[c,d]}=P_{[c,d]}P_{[a,b]},
\end{equation}
which are easily verified using definition (\ref{Pab}). Then we
see that operator $U_t$ is unitary, and satisfies
\begin{eqnarray}
\label{properties-of-U}
U_t^2=I,\quad
U_t^\dag=U_t,\quad
U_tU_s=U_sU_t,
\end{eqnarray}
where $I$ is the identity operator. The so-called
\textit{reflection process} $J_t\equiv J_t(U_t)$,
$t\in\mathbb{R}_+$, whose action on $\mathscr{E}$ is given by
\cite{Hudson86}
\bea
J_t|e(\g))=|e(U_t\g))
=\exp\lc U_t\int_0^\infty\d t'\,\g(t')b_{t'}^\dag\rc|),
\label{Jte}
\eea
inherits properties of the operator $U_t$ (\ref{properties-of-U}), i.e.
\begin{eqnarray}
\label{b12b}
J_t^2=I,\quad
J_t^\dag=J_t,\quad
J_tJ_s=J_sJ_t,
\end{eqnarray}
and does not change the vacuum: $(|J_t=(|$, $J_t|)=|)$.

Let us now consider new operators
\begin{eqnarray}
\mathfrak{b}_t=J_tb_t,\quad
\mathfrak{b}_t^\dag=b_t^\dag J_t.\label{JbbJ}
\end{eqnarray}
Apparently, they annihilate vacuums
\begin{eqnarray}
(|\mathfrak{b}_t^\dag=0,&\quad& \mathfrak{b}_t|)=0.
\end{eqnarray}
The following matrix elements
\begin{subequations}
\begin{align}
&(e(f)|[J_t,b_s]_{-\sigma}|e(\g))\nonumber\\
&\ =(e(f)|\{1-\sigma[\sigma_>+(\sigma_<-\sigma_>)
\theta(t-s)]\}\g(s)J_t|e(\g)),
\end{align}
\begin{equation}
\begin{split}
&(e(f)|[\mathfrak{b}_t,\mathfrak{b}_s^\dag]_{-\sigma}|e(\g))
=(e(f)|\delta(t-s)|e(\g))\\
&\ \phantom{{}={}}{}+(e(f)|J_tJ_s(\sigma_>\sigma_<-\sigma)f^*(s)\g(t)|e(\g)),
\end{split}
\end{equation}
\end{subequations}
are valid with arbitrary $f$ and $\g$,
where for arbitrary operators $A$ and $B$ the (anti-)commutator is
defined as
\begin{equation}
[A,B]_{-\sigma}=AB-\sigma BA,\;\sigma=\lc\!\!\!\!
\begin{array}{l}
\phantom{-}1~\text{for bosonic system,}\\
-1~\text{for fermionic system.}\\
\end{array}\right.
\end{equation}
Then the requirement of equal-time
(anti-)\-com\-mu\-ta\-ti\-vi\-ty between $J_t$ and $b_t$
\bea
[J_t,b_t]_{-\sigma}=0
\eea
gives $1-\sigma\sigma_<=0$,
while the requirement of canonical (anti-)commutation relation
\begin{eqnarray}
[\mathfrak{b}_t,\mathfrak{b}_s^\dag]_{-\sigma}=\delta(t-s)\label{ctcs}
\end{eqnarray}
leads to $\sigma_>\sigma_<-\sigma=0$.
All those conditions are satisfied when $\sigma_<=\sigma$ and $\sigma_>=+1$.
Then the operator $U_t$ turns out to be
\begin{eqnarray}
U_t=\sigma P_{[0,t]}+P_{(t,\infty)}.
\end{eqnarray}
Note that for a boson system, i.e.~$\sigma=1$, $U_t=I$ and the
operators $\mathfrak{b}_t$ and $\mathfrak{b}_t^\dag$ reduce,
respectively, to $b_t$ and $b_t^\dag$.

We see that the generalized quantum Brownian motion, defined by
\begin{eqnarray}
\mathfrak{B}_t=\int_0^t\d t'\,\mathfrak{b}_{t'},
\quad
\mathfrak{B}_t^\dag=\int_0^t\d t'\,\mathfrak{b}^\dag_{t'},
\label{b13}
\end{eqnarray}
with $\mathfrak{B}_0=0$, $\mathfrak{B}^\dag_0=0$, satisfies
\begin{eqnarray}
[\mathfrak{B}_t,\mathfrak{B}_s^\dag]_{-\sigma}=\min(t,s).
\end{eqnarray}
The case $\sigma=1$ represents the \textit{boson Brownian motion}
\cite{Hudson84a,Parthasarathy85}, whereas the case $\sigma=-1$ the
\textit{fermion Brownian motion} \cite{Hudson86}. Their increments
\begin{subequations}
\begin{align}
\d\mathfrak{B}_t&=\mathfrak{B}_{t+\d t}-\mathfrak{B}_t^{\phantom{\dag}}
=\mathfrak{b}_t\,\d t,\\
\d\mathfrak{B}_t^\dag&=\mathfrak{B}_{t+\d t}^\dag-\mathfrak{B}_t^\dag
=\mathfrak{b}_t^\dag\,\d t,
\end{align}
\end{subequations}
annihilate the vacuum, i.e.
\begin{eqnarray}
(|\d\mathfrak{B}^\dag_t=0,\quad\d\mathfrak{B}_t|)=0,
\end{eqnarray}
and their matrix elements read
\begin{subequations}
\begin{align}
(e(f)|\d\mathfrak{B}_t|e(\g))&=(e(f)|J_t\g(t)\d t|e(\g)),\\
(e(f)|\d\mathfrak{B}_t^\dag|e(\g))&=(e(f)|f^*(t)\d tJ_t|e(\g)),\quad\\
(e(f)|\d\mathfrak{B}_t\d\mathfrak{B}_t|e(\g))&=0,\\
(e(f)|\d\mathfrak{B}_t\d\mathfrak{B}_t^\dag|e(\g))&=\d t(e(f)|e(\g)).
\end{align}
\end{subequations}
Here we neglected the terms of higher order than $\d t$. The
latter equations are summarized in the following table of
multiplication rules for increments $\d\mathfrak{B}_t$ and
$\d\mathfrak{B}_t^\dag$:
\bea
\label{dBdBrule}
\ba{l|lll}
&\d\mathfrak{B}_t&\d\mathfrak{B}_t^\dag&\d t\\\hline
\d\mathfrak{B}_t&0&\d t&0\\
\d\mathfrak{B}_t^\dag&0&0&0\\
\d t&0&0&0\\
\ea
\eea

Quantum stochastic calculus with consideration for thermal degree
of freedom can be derived within the framework of Non-Equilibrium
Thermo Field Dynamics (NETFD). It is a unified formalism, which
enables us to treat dissipative quantum systems by the method
similar to the usual quantum mechanics and quantum field theory,
which accommodate the concept of the dual structure in the
interpretation of nature, i.e.~in terms of the operator algebra
and the representation space. Information about the general
structure of NETFD can be found in many papers and we refer first
of all to the original source \cite{Arimitsu85} and the review
article \cite{Arimitsu94}. With that in mind,
we consider now a tensor product space
$\hat\Gamma=\Gamma_s^0\otimes\tilde{\Gamma}_s^0$. Its vacuum
states $|)\!)$ and exponential vectors $|e(f,\g))\!)$ are defined
through the ``principle of correspondence'' \cite{Arimitsu85}
\begin{subequations}
\bea
|)\!)&\longleftrightarrow&|)(|,\\
|e(f,\g))\!)&\longleftrightarrow&|e(f))(e(\g)|.
\eea
\end{subequations}
Annihilation and creation operators acting on $\hat{\Gamma}$ are
defined through
\begin{subequations}
\bea
b_t|e(f,\g))\!)&\longleftrightarrow&b_t|e(f))(e(\g)|,\\
b_t^\dag|e(f,\g))\!)&\longleftrightarrow&b_t^\dag|e(f))(e(\g)|,\\
\tilde{b}_t|e(f,\g))\!)&\longleftrightarrow&|e(f))(e(\g)|b_t^\dag,\\
\tilde{b}_t^\dag|e(f,\g))\!)&\longleftrightarrow&|e(f))(e(\g)|b_t,
\eea
\end{subequations}
and similarly for $J_t$ and $\tilde{J}_t$, i.e.
\begin{subequations}
\bea
J_t|e(f,\g))\!)&\longleftrightarrow&J_t|e(f))(e(\g)|,\\
\tilde{J}_t|e(f,\g))\!)&\longleftrightarrow&|e(f))(e(\g)|J_t.
\eea
\end{subequations}
Algebra of commutation relations between these operators reads
\begin{align}
[b_t,b_s^\dag]&=[\tilde{b}_t,\tilde{b}_s^\dag]=\delta(t-s),
&&[b_t,\tilde{b}_s]=[b_t,\tilde{b}_s^\dag]=0,\\\relax
[J_t,\tilde{b}_s]&=[\tilde{J}_t,b_s]=0,
&&[J_t,b_t]_{-\sigma}=[\tilde{J}_t,\tilde{b}_t]_{-\sigma}=0.\nonumber
\end{align}

Let us now consider new operators defined by
\begin{subequations}
\label{no-tilde-and-tilde-fbt}
\begin{align}
\mathfrak{b}_t&=J_tb_t,&\mathfrak{b}_t^\dag&=b_t^\dag J_t,\label{no-tilde-fb}\\
\tilde{\mathfrak{b}}_t&=\hat\tau\tilde{J}_t\tilde{b}_t,&
\tilde{\mathfrak{b}}_t^\dag&=\hat\tau\tilde{b}_t^\dag\tilde{J}_t,
\label{tilde-fbt}
\end{align}
\end{subequations}
where $\hat\tau$ is an operator introduced to ensure the following relations:
\begin{subequations}
\label{req-for-fbt}
\bea
[\mathfrak{b}_t,\mathfrak{b}_s^\dag]_{-\sigma}&=&
[\tilde{\mathfrak{b}}_t,\tilde{\mathfrak{b}}_s^\dag]_{-\sigma}=\delta(t-s),
\label{req-fbt}\\\relax
[\mathfrak{b}_t,\tilde{\mathfrak{b}}_s]_{-\sigma}&=&
[\mathfrak{b}_t,\tilde{\mathfrak{b}}_s^\dag]_{-\sigma}=0.
\label{tau-condition-1}
\eea
\end{subequations}
In order to do that it should commute with $J_t$ and
$\tilde{J}_t$, and (anti-)\-com\-mu\-te with $b_t$ and $\tilde{b}_t$:
\begin{subequations}
\bea
[\hat\tau,J_t]&=&[\hat\tau,\tilde{J}_t]=0,\\\relax
[\hat\tau,b_t]_{-\sigma}&=&[\hat\tau,\tilde{b}_t]_{-\sigma}=0.
\eea
\end{subequations}
Other properties of operator $\hat\tau$ can be derived using definitions
(\ref{no-tilde-and-tilde-fbt}) and requirements (\ref{req-for-fbt}). For
example, from (\ref{tilde-fbt}) and second equality in (\ref{req-fbt}) we have
\bea
[\tilde{\mathfrak{b}}_t,\tilde{\mathfrak{b}}_s^\dag]_{-\sigma}=
\sigma\hat\tau^2[\tilde{J}_t\tilde{b}_t,\tilde{b}_s^\dag\tilde{J}_s]_{-\sigma}=
\sigma\hat\tau^2\delta(t-s),
\eea
which gives $\hat\tau^2=\sigma$.
On the other hand, commutativity of tilde conjugation
and hermitian conjugation, i.e.~$(\tilde{\mathfrak{b}}_t)^\dag=\tilde{b}_t^\dag\tilde{J}_t\hat\tau^\dag
=\sigma\hat\tau^\dag\tilde{b}_t^\dag\tilde{J}_t$, means
$\hat\tau^\dag=\sigma\hat\tau$.
From the requirement that
$(\tilde{\mathfrak{b}}_t)^\sim=\mathfrak{b}_t$ and noting that
$(\tilde{\mathfrak{b}}_t)^\sim=\hat\tau(\hat\tau)^\sim J_tb_t
=\hat\tau(\hat\tau)^\sim\mathfrak{b}_t$, one obtains
$(\hat\tau)^\sim=\hat\tau^\dag$.

Thermal degree of freedom can be introduced by the Bogoliubov
transformation in $\hat{\Gamma}$. For this purpose we re\-qui\-re that
the expectation value of $\mathfrak{b}_t^\dag\mathfrak{b}_s$
should be
\begin{eqnarray}
\la\mathfrak{b}_t^\dag\mathfrak{b}_s\ra=\bar{n}\delta(t-s)\label{req}
\end{eqnarray}
with $\bar{n}\in\mathbb{R}_+$, where $\la\ldots\ra=\la|\ldots|\ra$
indicates the expectation with respect to tilde invariant thermal
vacuums $\la|$ and $|\ra$. The requirement
(\ref{req}) is consistent with the thermal state conditions for states
$\la|$ and $|\ra$ such that
\begin{eqnarray}
\la|\tilde{\mathfrak{b}}_t^\dag=\tau^*\la|\mathfrak{b}_t,\quad
\tilde{\mathfrak{b}}_t|\ra=\frac{\tau\bar{n}}{1+\sigma\bar{n}}\mathfrak{b}_t^\dag|\ra,
\label{TSC-b}
\end{eqnarray}
$\tau=1$ for bosonic system and $\tau=i$ for fermionic one.

Let us introduce annihilation and creation operators
\begin{equation}
\label{aco-c}
\mathfrak{c}_t=[1+\sigma\bar{n}]\mathfrak{b}_t-\sigma\tau\bar{n}\tilde{\mathfrak{b}}_t^\dag,\quad
\tilde{\mathfrak{c}}_t^{\venus}=\tilde{\mathfrak{b}}_t^\dag-\sigma\tau\mathfrak{b}_t,
\end{equation}
and their tilde conjugates. From (\ref{TSC-b}) one has
\begin{eqnarray}
\la|\mathfrak{c}_t^{\venus}=\la|\tilde{\mathfrak{c}}_t^{\venus}=0,\quad
\mathfrak{c}_t|\ra=\tilde{\mathfrak{c}}_t|\ra=0.
\end{eqnarray}
With the thermal doublet notations
\begin{align}
\bar{\mathfrak{b}}_t^\mu&=\lp\mathfrak{b}_t^\dag,-\tau\tilde{\mathfrak{b}}_t\rp,
&\mathfrak{b}_t^\nu&=\mathrm{column}\lp\mathfrak{b}_t,
\tau\tilde{\mathfrak{b}}_t^\dag\rp,\\
\bar{\mathfrak{c}}_t^\mu&=\lp\mathfrak{c}_t^{\venus},-\tau\tilde{\mathfrak{c}}_t\rp,
&\mathfrak{c}_t^\nu&=\mathrm{column}\lp\mathfrak{c}_t,
\tau\tilde{\mathfrak{c}}_t^{\venus}\rp,
\end{align}
(\ref{aco-c}) and their tilde conjugates can be written in form of
the Bogoliubov transformation
\begin{eqnarray}
\mathfrak{c}_t^\mu=B^{\mu\nu}\mathfrak{b}_t^\nu,\quad
\bar{\mathfrak{c}}_t^\nu=\bar{\mathfrak{b}}_t^\mu[B^{-1}]^{\mu\nu},
\end{eqnarray}
\begin{eqnarray}
\text{with}\quad\quad B^{\mu\nu}=\lp
\ba{rr}
1+\sigma\bar{n}&-\sigma\bar{n}\\
-1&1\\
\ea\rp.
\end{eqnarray}
This transformation is canonical since new operators satisfy
canonical (anti-)commutation relations
\bea
[\mathfrak{c}_t,\mathfrak{c}_s^{\venus}]_{-\sigma}=\delta(t-s).
\eea

In the following, we will use the representation space constructed
on vacuums $\la|$ and $|\ra$. Let $\Gamma^\beta$ denotes the Fock
space spanned by the basic bra- and ket-vectors introduced by a
cyclic operations of $\mathfrak{c}_t$, $\tilde{\mathfrak{c}}_t$ on
the thermal bra-vacuum $\la|$, and of $\mathfrak{c}_t^{\venus}$,
$\tilde{\mathfrak{c}}_t^{\venus}$ on the thermal
ket-va\-cu\-um~$|\ra$.
Quantum Brownian motion at finite temperature is defined in the
Fock space $\Gamma^\beta$ by operators
\begin{eqnarray}
\mathfrak{B}_t^\sharp=\int_0^t\d s\,\mathfrak{b}_s^\sharp,\quad
\tilde{\mathfrak{B}}_t^\sharp=\int_0^t\d s\,\tilde{\mathfrak{b}}_s^\sharp,
\end{eqnarray}
with $\mathfrak{B}_0^\sharp=0$ and
$\tilde{\mathfrak{B}}_0^\sharp=0$, where $\sharp$ stands for
null or dagger. The explicit representation of processes
$\mathfrak{B}_t^\sharp$ and $\tilde{\mathfrak{B}}_t^\sharp$ can be
performed in terms of the Bogoliubov transformation. The couple
$\mathfrak{B}_t$ and $\mathfrak{B}_t^\dag$, for example, is
calculated as
\begin{subequations}
\setlength{\arraycolsep}{1pt}
\begin{align}
\mathfrak{B}_t&=\int_0^t\d s\Big[\mathfrak{c}_s
+\sigma\tau\bar{n}\tilde{\mathfrak{c}}_s^{\venus}\Big]
=\mathfrak{C}_t+\sigma\tau\bar{n}\tilde{\mathfrak{C}}_t^{\venus},\\
\mathfrak{B}_t^\dag&=\int_0^t\d s
\Big[[1+\sigma\bar{n}]\mathfrak{c}_s^{\venus}+\tau\tilde{\mathfrak{c}}_s\Big]
=[1+\sigma\bar{n}]\mathfrak{C}_t^{\venus}+\tau\tilde{\mathfrak{C}}_t,
\end{align}
\end{subequations}
where we defined new operators
\begin{eqnarray}
\mathfrak{C}_t^\sharp=\int_0^t\d s\,\mathfrak{c}_s^\sharp,\quad
\tilde{\mathfrak{C}}_t^\sharp=\int_0^t\d s\,\tilde{\mathfrak{c}}_s^\sharp,
\end{eqnarray}
with $\mathfrak{C}_0^\sharp=0$ and
$\tilde{\mathfrak{C}}_0^\sharp=0$, and $\sharp$ standing for null
or the Ve\-nus-mark.
Since matrix elements of $\d\mathfrak{C}_t^\sharp$ and
$\d\tilde{\mathfrak{C}}_t^\sharp$ in thermal space $\Gamma^\beta$
read
\begin{subequations}
\begin{align}
\la\d\mathfrak{C}_t\ra=\la\d\tilde{\mathfrak{C}}_t\ra&=
\la\d\mathfrak{C}_t^{\venus}\ra=\la\d\tilde{\mathfrak{C}}_t^{\venus}\ra=0,\\
\la\d\mathfrak{C}_t^{\venus}\d\mathfrak{C}_t\ra&=
\la\d\tilde{\mathfrak{C}}_t\d\tilde{\mathfrak{C}}_t^{\venus}\ra=0,\\
\la\d\mathfrak{C}_t\d\mathfrak{C}_t^{\venus}\ra&=
\la\d\tilde{\mathfrak{C}}_t^{\venus}\d\tilde{\mathfrak{C}}_t\ra=\d t,
\end{align}
\end{subequations}
calculation of moments of quantum Brownian motion in the thermal
space $\Gamma^\beta$ can be performed, for instance, as
{\setlength{\arraycolsep}{2pt}
\begin{align}
\la\d\mathfrak{B}_t\d\mathfrak{B}_t^\dag\ra&=\La\!\lp\d\mathfrak{C}_t
+\sigma\tau\bar{n}\d\tilde{\mathfrak{C}}_t^{\venus}\rp
\!\!\lp[1+\sigma\bar{n}]\d\mathfrak{C}_t^{\venus}
+\tau\d\tilde{\mathfrak{C}}_t\rp\!\Ra\nonumber\\
&=[1+\sigma\bar{n}]\la\d\mathfrak{C}_t\d\mathfrak{C}_t^{\venus}\ra
=[1+\sigma\bar{n}]\d t.
\end{align}\relax}
Repeating this for other pair products of
$\d\mathfrak{B}_t^\sharp$, $\d\tilde{\mathfrak{B}}_t^\sharp$ and
$\d t$, multiplication rules for these increments can be
summarized in the following table:
{\setlength{\arraycolsep}{3pt}
\begin{eqnarray}\label{st-calc}
\begin{array}{l|ccccc}
&\d\mathfrak{B}_t&\d\mathfrak{B}_t^\dag&\d\tilde{\mathfrak{B}}_t&\d\tilde{\mathfrak{B}}_t^\dag&\d t\\\hline
\d\mathfrak{B}_t&0&\phantom{\sigma\tau}[1+\sigma\bar{n}]\d t&\tau\bar{n}\d t&0&0\\
\d\mathfrak{B}_t^\dag&\phantom{\sigma\tau}\bar{n}\d t&0&0&\tau[1+\sigma\bar{n}]\d t&0\\
\d\tilde{\mathfrak{B}}_t&\sigma\tau\bar{n}\d t&0&0&\phantom{\tau}[1+\sigma\bar{n}]\d t&0\\
\d\tilde{\mathfrak{B}}_t^\dag&0&\sigma\tau[1+\sigma\bar{n}]\d t&\phantom{\tau}\bar{n}\d t&0&0\\
\d t&0&0&0&0&0\\
\end{array}\quad
\end{eqnarray}}\relax
So far we know, result (\ref{st-calc}) is the first report about
\textit{quantum boson and fermion stochastic calculus} valid for the
stationary non-equilibrium case. Similar result for purely boson
system was obtained in ref. \citen{Saito97}. In ref. \citen{Kobryn03} we
derive stochastic calculus valid for both stationary and
non-stationary cases and use them for analysis of quantum
stochastic differential equations. The obtained equations include
quantum Langevin equation and quantum stochastic Liouville
equation together with the corresponding master equation. The
Fokker--Planck equation is derived by taking the random average of
the corresponding stochastic Liouville equation. The relation
between the Langevin equation and the stochastic Liouville
equation, as well as between the Heisenberg equation for operators
of gross variables and the Fokker--Planck equation obtained there,
is similar to one between the Heisenberg equation and the
Schr\"o\-din\-ger equation in quantum mechanics and field theory.
Application of quantum stochastic differential equations for
particular problems will be presented elsewhere.


\end{document}